\documentclass{emulateapj}
\usepackage{epsfig}
\usepackage{graphicx}
\usepackage{color}
\newcommand{\msun}{\,\rm M_\odot}
\newcommand{\VLII}{{\it Via Lactea II}}
\newcommand{\etal}{et al.\ }

\newcommand{\be}{\begin{equation}}
\newcommand{\ee}{\end{equation}}
\newcommand{\ba}{\begin{eqnarray}}
\newcommand{\ea}{\end{eqnarray}}

\newcommand{\rtwo}{r_{200}}

\def\spose#1{\hbox to 0pt{#1\hss}}
\def\lta{\mathrel{\spose{\lower 3pt\hbox{$\mathchar"218$}} \raise 2.0pt\hbox{$\mathchar"13C$}}}
\def\gta{\mathrel{\spose{\lower 3pt\hbox{$\mathchar"218$}} \raise 2.0pt\hbox{$\mathchar"13E$}}}

\lefthead{Madau \etal}
\righthead{Fossil remnants of reionization}



\begin{document}
\submitted{ApJL, in press}

\title{Fossil remnants of reionization in the halo of the Milky Way}

\author{P. Madau\altaffilmark{1}, M. Kuhlen\altaffilmark{2}, J. Diemand\altaffilmark{1},
B. Moore\altaffilmark{3}, M. Zemp\altaffilmark{1,4}, D. Potter\altaffilmark{3}, \& J. 
Stadel\altaffilmark{3}} 
\altaffiltext{1}{Department of Astronomy \& Astrophysics, University of California, 
Santa Cruz, CA 95064, pmadau,diemand@ucolick.org.}
\altaffiltext{2}{Institute for Advanced Study, Einstein Drive, Princeton, NJ 08540, mqk@ias.edu.}
\altaffiltext{3}{Institute for Theoretical Physics, University Zurich, 
Winterthurerstrasse 190, 8057 Zurich, Switzerland, dpotter,moore,stadel@physik.uzh.ch.}
\altaffiltext{4}{Astronomy Department, University of Michigan, Ann Arbor, MI, 48109,
mzemp@umich.edu.}

\begin{abstract}
  Our recently completed one billion particle \VLII\ simulation of a
  Milky Way-sized dark matter halo resolves over 50,000
  gravitationally bound clumps orbiting today within the virialized
  region of the main host. About 2,300 of these subhalos
  have one or more ``progenitors'' with $M>10^6\,\msun$ at
  redshift $z=11$, i.e. massive enough for their gas to have cooled via
  excitation of H$_2$ and fragmented prior to the epoch of cosmic
  reionization. We count 4,500 such progenitors: if these were able 
  to convert a fraction of their gas
  content into very metal-poor stars with a Salpeter initial mass function
  (IMF), they would be shining today
  with a visual magnitude $M_V=6.7$ per solar mass in stars. Assuming
  a universal baryon fraction, we show
  that mean star formation efficiencies as low as 0.1\% 
  in progenitors $\ll 10^8\,\msun$ would
  overproduce the abundance of the faint Galatic dwarf spheroidals
  observed by the Sloan Digital Sky Survey. 
  Star formation at first light must have occurred either with an IMF
  lacking stars below $0.9\,\msun$, or was intrinsically very
  inefficient in small dark matter halos. If the latter, our results 
  may be viewed as another hint of a minimum scale in galaxy formation.
\end{abstract}

\section{Introduction}

Galaxy halos are one of the crucial testing grounds for structure
formation scenarios, as they contain the imprints of past accretion
events, from before the epoch of reionization to the present. In the
standard $\Lambda$CDM concordance cosmology, objects like the halo of
our Milky Way are assembled via the hierarchical merging and accretion
of many smaller progenitors. Subunits collapse at high redshift, are
dense, and have cuspy density profiles, and when they merge into
larger hosts they are able to resist tidal disruption. Indeed, the
{\it Via Lactea Project}, a suite of some of the largest cosmological
simulations of Galactic dark matter substructure, has shown that
galaxy halos today are teeming with surviving
``subhalos'' (Diemand \etal 2007, 2008). 
The predicted subhalo counts vastly exceed the number of known
satellites of the Milky Way, a ``substructure problem'' that has been
the subject of many recent studies. While a full characterization of this
discrepancy is hampered by luminosity bias in the observed satellite
luminosity function (LF) (Koposov \etal 2008), it is
generally agreed that cosmic reionization may offer a plausible
solution to the apparent conflict between the Galaxy's relatively
smooth stellar halo and the extremely clumpy cold dark matter distribution. 
In this hypothesis, photoionization heating
{\it after reionization breakthrough} reduces the star forming ability
of newly forming halos that are not sufficiently massive to accrete
intergalactic gas (e.g. Bullock \etal 2000; Somerville 2002; Kravtsov
\etal 2004; Ricotti \& Gnedin 2005; Moore \etal 2006; Strigari \etal
2007; Simon \& Geha 2007; Madau \etal 2008). The exact value of this
mass threshold remains uncertain, as it depends on poorly known
quantities like the amplitude and spectrum of the ionizing background
and the ability of the system to self-shield against UV radiation
(e.g. Efstathiou 1992; Dijkstra \etal 2004).

In this {\it Letter} we take a different approach and focus instead on
the sources of reionization itself. We use our one billion particle
\VLII\ simulation to constrain the character of star formation
(initial mass function and efficiency of gas conversion into stars) at
first light, i.e. in subgalactic halos {\it prior to the epoch of
reionization}. The starting point of our investigation is the
finding that as many as 2,300 bound clumps that survive today in 
the \VLII\ halo have early progenitors that were massive
enough for their gas to cool via excitation of molecular hydrogen and
fragment before reionization breakthrough.  If low-mass stars formed
in the process, their hosts would be shining today as faint Milky Way
satellites.

\section{Via Lactea II}

Our recently completed \VLII\ simulation, one of the highest-precision
calculation of the assembly of the Galactic CDM halo to date (Diemand
\etal 2008), offers the best opportunity for a systematic
investigation of the fossil records of reionization in the halo of the
Milky Way.  \VLII\ employs just over one billion $4,100 \msun$
particles to model the formation of a $M_{200}=1.9\times
10^{12}\,\msun$ Milky-Way size halo and its substructure. It resolves
50,000 subhalos today within the host's $\rtwo=402$ kpc (the
radius enclosing an average density 200 times the mean matter value).

\begin{figure}[thb]
\begin{center}
\includegraphics*[width=0.45\textwidth]{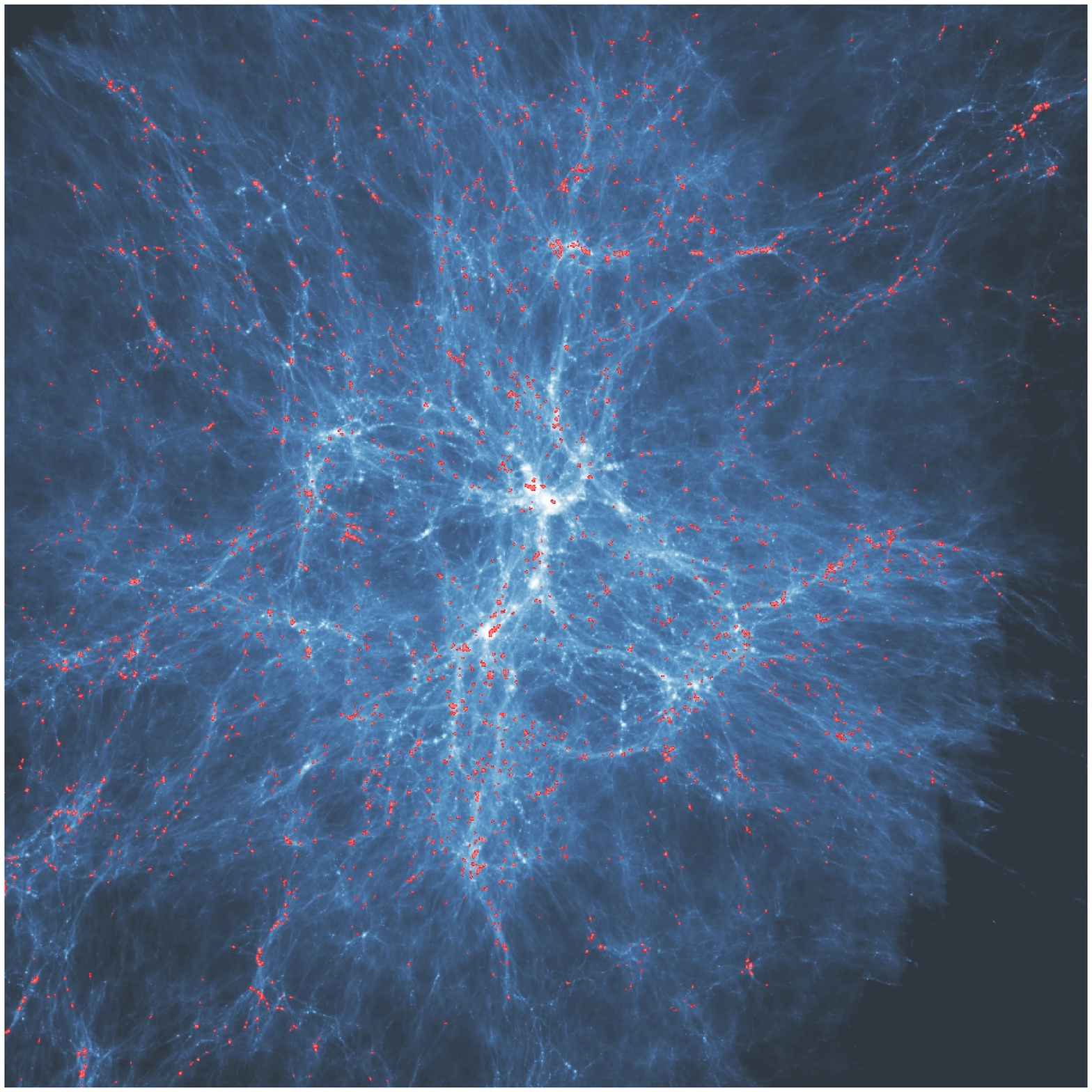}
\includegraphics*[width=0.45\textwidth]{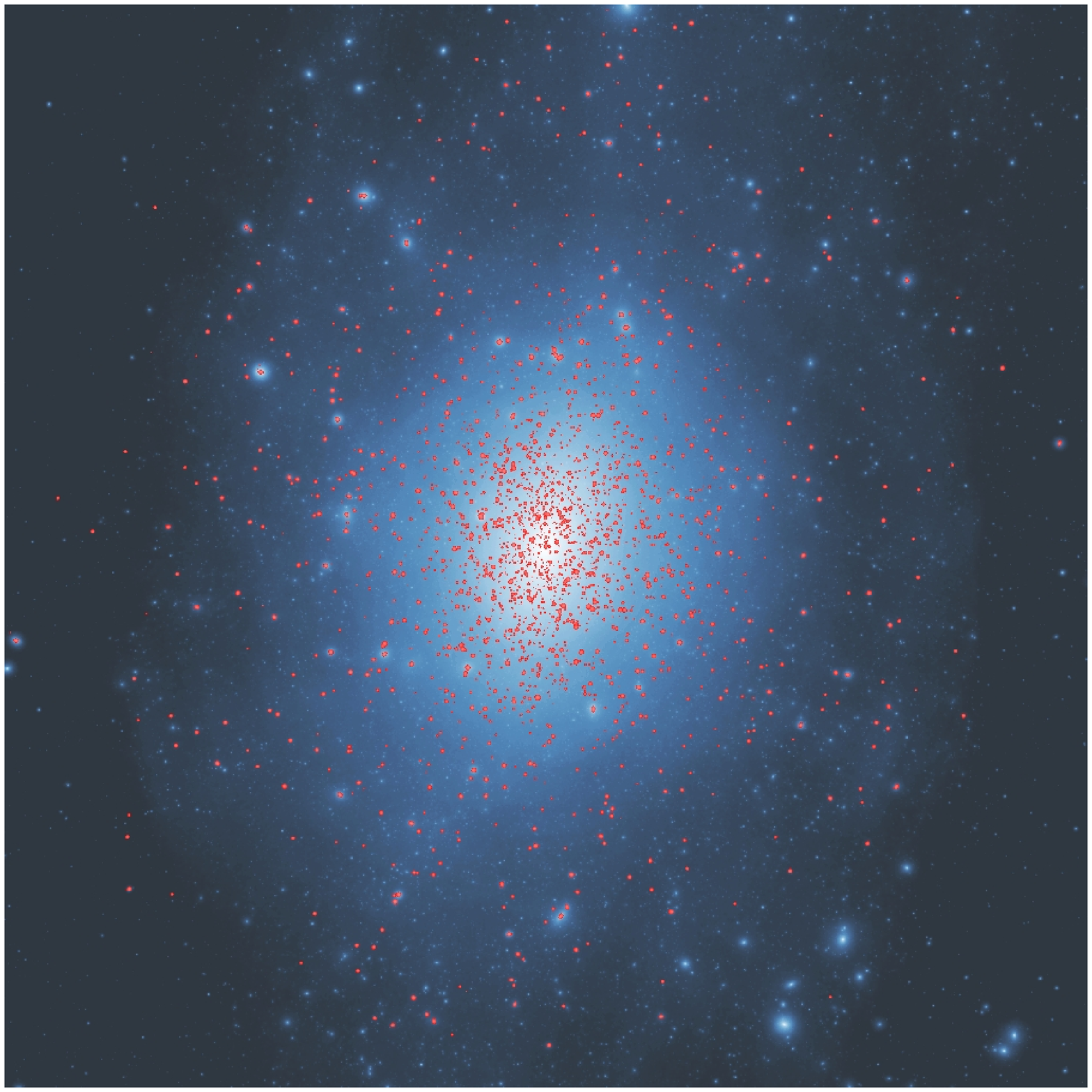}
\end{center}
\caption{ {\it Top:} Projected dark matter density map of \VLII\
at $z=11$. The image covers 3.5 comoving Mpc across and shows (magenta) all 
progenitor minihalos with $M>10^6\,\msun$ that have a bound descendant 
today within $\rtwo$. {\it Bottom:} the $z=0$ descendants of those early 
progenitors. The image covers 800 kpc across.   
}
\label{fig1a}
\vspace{-0.3cm}
\end{figure}

\begin{figure}[thb]
\vspace{-0.7cm}
\centering
\includegraphics*[width=0.49\textwidth]{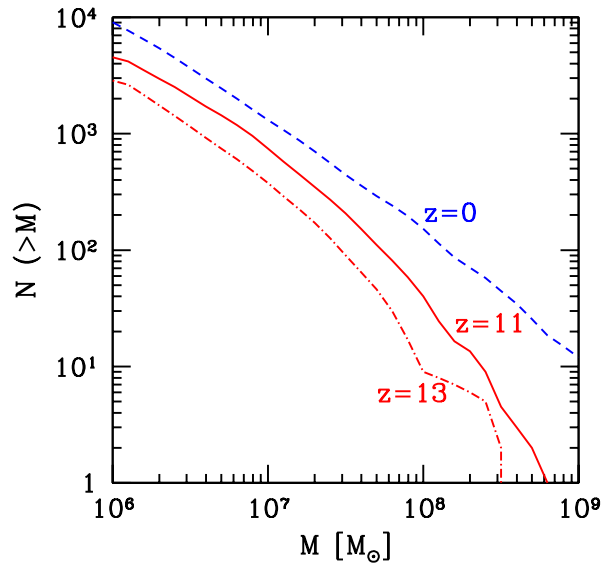}
\caption{ \VLII\ cumulative substructure mass function at
  different epochs.  {\it Solid curve:} All 4,500 progenitor halos
  with $M>10^6\,\msun$ at $z=11$ that contribute to a
  gravitationally bound descendant (2,300 of them) today within
  $\rtwo$. {\it Dot-dashed curve:} same for the 2,900
  $M>10^6\,\msun$ progenitors at $z=13$. Also shown for comparison
  is the mass function of all \VLII\ subhalos at the present epoch
  ({\it dashed curve}).  }
\label{fig1}
\vspace{-0.3cm}
\end{figure}

The {\it Wilkinson Microwave Anisotropy Probe (WMAP)} 5-year data
require the universe to be fully reionized by redshift $z=11.0\pm 1.4$
(Dunkley \etal 2008).  Assuming that the region around the Milky Way
was reionized at about the same epoch of the universe as a whole, we
can trace the progenitors of present-day surviving substructure back
to a time {\it prior to reionization breakthrough}, i.e. before star
formation was quenched by photoheating. We use the 6DFOF group finder 
described in Diemand \etal (2007) to identify peaks in phase-space
density at $z=11$ and $z=0$. The resulting groups contain between 16
and a few thousand particles linked together in the centers of halos
and subhalos. Note that the gravitationally bound region often extends
beyond this central group. We link a descendant $z=0$ group ``A'' to a
high redshift progenitor ``B'' if ``A'' contains more than 10
particles from ``B'' and more than any other descendant of ``B''. Thus
a descendant can have more than one progenitor, but a progenitor is
linked to at most one descendant. Using the halo centers provided by
6DFOF ensures that material from ``B'' does indeed contribute to the
central regions of ``A'', i.e. that a star cluster that formed in the
core of a progenitor clump would end up today in the dwarf galaxy of
a descendant subhalo.
We include in our analysis only progenitors that either survive
individually during the hierarchical clustering process or contribute
to a gravitationally-bound descendant at $z=0$. About 10,000 ``first generation''
systems above $10^6\,\msun$ are totally
disrupted by tidal forces: for half of them all of their particles lie
today within $\rtwo$ and contribute to the smooth stellar halo.
Figure \ref{fig1a} shows an image of all \VLII\ progenitor halos 
above $10^6\,\msun$ at $z=11$ with a bound descendants within $\rtwo$ today,
while Figure \ref{fig1} shows the cumulative mass functions of 
progenitors and parents. We track forward in time
4,500 of these early minihalos into 2,300 descendants
today.\footnote{Note that the size of the
  Galaxy halo is not known very precisely, since the effects of disk
  formation on the total mass distribution are still poorly
  constrained. When comparing \VLII\ to the Milky Way, this
  translates into an uncertainty in the abundance of subhalos of about
  a factor of two (Klypin \etal 2002; Dutton \etal 2007).}
\section{Star formation at first light}

Cosmological hydrodynamics simulations of structure formation in
a $\Lambda$CDM universe have found that in the early collapse of $\gta 
10^6\,\msun$ systems, enough H$_2$ is produced to cool the gas 
and allow star formation within a Hubble time (e.g. Abel \etal 2002; 
Bromm \etal 2002). Figure \ref{fig1} shows that a few hundred progenitors 
are above the ``atomic cooling'' mass threshold of $\sim 3\times 10^7\,
\msun$ ($T_{\rm vir}\gta 10^4\,$K), where gas can cool and fragment via excitation
of hydrogen Ly$\alpha$. Numerical studies also suggest that, while
primordial stars were likely very massive and formed in isolation at
$z\gta 20$ (e.g. O'Shea \& Norman 2007), low-mass second-generation
stars can form as soon as a minimum pre-enrichment level of
$Z=10^{-5\pm 1}\,Z_\odot$ is reached (e.g.  Schneider \etal 2002).
The existence of 0.8 $\msun$ extremely iron-deficient stars in the
halo of the Milky Way (Christlieb \etal 2002; Frebel \etal 2005)
indicates that low-mass star formation was possible at very-low
metallicities (e.g. Tumlinson 2007).
\begin{figure}[thb]
\vspace{-0.7cm}
\centering
\includegraphics*[width=0.49\textwidth]{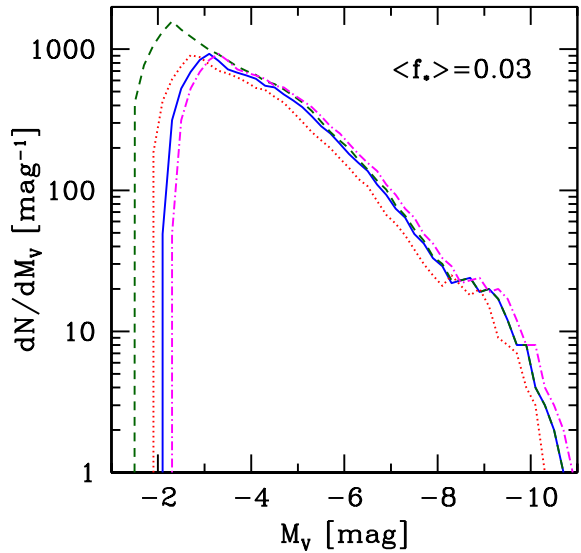}
\caption{The luminous fossil remnants of reionization. {\it Solid curve:}
present-day LF of all \VLII\ subhalos
that had one or more progenitor with $M>10^6\,\msun$ at $z=11$. At this epoch,
a fraction $f_*$ of the gas content of each progenitor is turned
instantaneously into stars with a Salpeter IMF and
metallicity $Z=Z_\odot/200$. {\it Dashed curve:} same for $M>5\times 10^5\,\msun$.
{\it Dot-dashed curve:} same for a Chabrier (2003) IMF.
{\it Dotted curve:} same for $Z=Z_\odot/5$.
In all the above cases visual magnitudes at age 13.7 Gyr have been calculated using
Bruzual \& Charlot's (2003) models (1994 Padova tracks).
}
\vspace{0.0cm}
\label{fig2}
\end{figure}
In order to constrain the number of low-mass stars that lit up dark
matter halos before reionization, our default model makes a number of simplifying 
assumptions: 1) all progenitor hosts above
$M(z=11)=10^6\,\msun$ have a gas content given by the universal
baryon fraction, $M_{\rm gas}=(\Omega_b/\Omega_m)M=0.17\,M$; 2)
at $z\gta 11$ a fraction $f_*$ of this gas is turned into very
metal-poor stars with $Z=Z_\odot/200$ (the lowest metallicity 
allowed by Bruzual \& Charlot's 2003 stellar population synthesis models)
and a standard Salpeter initial mass function (IMF) in the range $0.1<m_*<100\,
\msun$. The star formation efficiency is independent of $M$; 3) star formation is
suppressed at later epochs in all progenitors and their descendants;
4) primordial stellar
systems are deeply embedded in progenitor minihalos and remain largely
unaffected by tidal stripping even if their hosts are not. The
complete tidal disruption of a host, however, also destroys its
stellar system.

We are now in the position to construct the present-day LF of 
the fossil remnants of reionization in the halo of the
Milky Way. According to Bruzual \&
Charlot (2003), a stellar population undergoing an instantaneous burst
will be shining at age 13.7 Gyr with a visual magnitude per solar mass
in stars of 6.7 mag (Salpeter IMF, $Z=Z_\odot/200$). The steep
LF of \VLII\ first-generation halos predicted in
this case is shown in Figure \ref{fig2} for $\langle f_*\rangle =0.03$.  Luminous
remnants are distributed between $M_V=-2$ and $M_V=-10$, and there are
more than two hundred relatively bright objects with $M_V<-6$, i.e. as
bright as Bo\"{o}tes. A mean star formation
efficiency three times smaller would shift this curve 1.2 mag to the left.
The cutoff at faint magnitudes simply reflects our assumption that
clumps below $M(z=11)=10^6\, \msun$ never form stars: the effect
of lowering this mass threshold by a factor of two is also shown for
comparison. Note that implausibly increasing the metallicity of early
stars to $Z=Z_\odot/5$ would only shift the curve 0.3 mag to the left
(Bruzual \& Charlot 2003).  Similarly, a Chabrier (2003) IMF in the
same mass range instead of Salpeter has only a small (0.2 mag)
brightening effect.

\section{Constraints from the SDSS}

\begin{figure}[thb]
\vspace{-0.7cm}
\centering
\includegraphics*[width=0.49\textwidth]{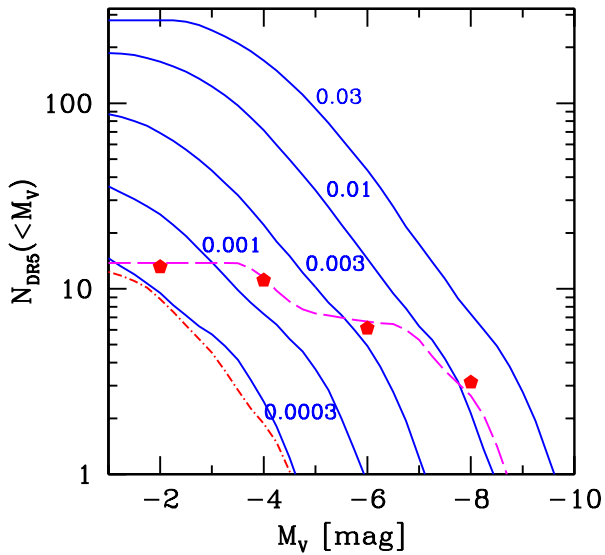}
\caption{ \VLII\ first-generation ($z=11$) subhalos detectable today by
the SDSS DR5. All curves assume $M>10^6\,\msun$, a Salpeter IMF, and $Z=Z_\odot/200$.
The data points are taken from the compilation of Tollerud \etal (2008).
{\it Labelled solid curves:} $\langle f_*\rangle =0.03,0.01,0.003,0.001,0.0003$.
{\it Dashed curve:} mass-dependent star formation efficiency, $\langle f_*\rangle
=(0.02,0.0025,0)$
for $M$ in the range $(>7\times 10^7,3.5\times 10^7$-$7\times 10^7,<3.5\times
10^7\,\msun)$.
Also plotted for comparison is the LF of the remnants of $z=13$ halos with
$\langle f_*\rangle=0.0005$ ({\it short dot-dashed curve} at the bottom).
}
\vspace{0.0cm}
\label{fig3}
\end{figure}

Over the last two years, the Sloan Digital Sky Survey (SDSS) has doubled 
the number of known Milky Way dwarf spheroidals (dSphs) brighter than 
$M_V=-2$ (e.g. Belokurov \etal 2007). 
An accurate estimate of the total number of dwarfs 
requires a correction to the observed LF that accounts for completeness limits 
and that depends on the unknown intrinsic spatial distribution of sources. Rather than 
producing an ``unbiased'' list of Galactic satellites as in
Tollerud \etal (2008), we apply here the completeness limit of the SDSS DR5 
to our predicted fossil remnants in \VLII\ (Fig. \ref{fig2}) to construct an 
artificial DR5 sample of luminous first-generation substructure. As computed by 
Koposov \etal (2008) (and fitted by Tollerud \etal 2008), the maximum accessible distance 
beyond which an object of magnitude $M_V$ will go undetected by the SDSS is
\begin{equation}
r_{\rm max}=\left({3\over 4\pi f_{\rm DR5}}\right)^{1/3}\,10^{(-0.6M_V-5.23)/3}\,{\rm Mpc}, 
\end{equation}
where $f_{\rm DR5}=0.194$ is the fraction of the sky covered by DR5.
Therefore, only subhalos brighter than $M_V=-6.6$ will be detected all
the way to $\rtwo$: first-generation remnants with $M_V=-4$ will be included
in the sample only if they are closer than 120 kpc. (The above
detection limit applies to objects with surface brightness $\lta
30$ mag arcsec$^{-2}$, i.e. we are implicitly assuming that our luminous
progenitors are less diffuse than this.) As a first order
correction we also assume that subhalos
within $r_{\rm max}$ are detected with 100\% efficiency, and account
for the partial sky coverage of SDSS by multiplying the number of
detectable subhalos by $f_{\rm DR5}$. The resulting cumulative LF $N_{\rm
DR5}(<M_V)$ is plotted in Figure \ref{fig3} for values of 
$\langle f_*\rangle$ that decrease from 3\% down to $3\times 
10^{-4}$. In this range, the faint-end of the predicted LF is dominated
by the remnants of $10^6-10^7\,\msun$ progenitors.  
For an efficiency of 1\%, SDSS should have detected $\sim 150$ first-generation systems 
brighter than $M_V=-2$, while only a dozen are actually observed.
The known Milky Way satellites are also plotted for
comparison: these include the SDSS eleven dwarfs with $M_V<-2$
(including Leo T which, at a distance of 417 kpc, lies just outside
$\rtwo$) as well as the eleven (times $f_{\rm DR5}$ to correct for the
partial sky coverage) classical (pre-SDSS) bright satellites. 
{\it Clearly, mean star formation efficiencies as low as 0.1\% in progenitors 
below $10^7\,\msun$ would overproduce the abundance of ultra-faint dSphs.} 
The mean efficiency of star formation must actually scale with halo mass 
in order to reproduce the observations with first-generation remnants. This is because a
constant efficiency translates into a steep LF that
follows the steep mass function of CDM field halos. By contrast, the 23 
known satellites of the Milky Way have a flat LF
and shine with luminosities ranging from about a thousand to a billion times 
solar. The dashed curve in Figure \ref{fig3} shows a toy model with a mass-dependent 
efficiency, $\langle f_*\rangle =(0.02,0.0025,0)$ for $M$ in the range 
$M=(>7\times 10^7,3.5\times 10^7$-$7\times 10^7,<3.5\times 10^7\,\msun)$.  
Note how the data allow efficiencies of order 1\% only above the atomic cooling 
threshold.
%
 
\section{Summary}

In this work we have used results from the one billion
particle \VLII\ simulation to study for the first time the fossil
signatures of the pre-reionization epoch in the Galactic halo and set
constraints on the baryonic building blocks of today's galaxies. 
We have traced the progenitors of
present-day surviving substructure back to a time {\it prior to 
reionization breakthrough}, i.e. before star formation was quenched
by an external UV background. We have then populated early-forming minihalos 
with very metal-poor stars following a Salpeter IMF, and looked at the
photometric properties today of their 2,300 descendants within
$\rtwo$. We have shown that even star 
formation efficiencies as low as 0.1\% in progenitor minihalos
below $10^7\,\msun$ would overproduce the abundance of Galatic dSphs observed by the
SDSS. This value should be regarded as an upper limit since luminous 
dSphs are known to have formed a fraction of their stars after reionization 
(e.g. Orban \etal 2008). It is also meant to be taken as a mean value,
since an alternative possibility is to allow stars to form efficiently in a small 
fraction of minihalos at some given mass and to suppress 
star formation in the remainders, rather than reducing the efficiency 
across the board. Note that the parameter $f_*$ is degenerate with 
the baryon fraction: if the gas content of early virialized structure 
was smaller than the universal value
(as found, e.g., by O'Shea \& Norman 2007), then our limits on the 
star formation efficiency would be correspondingly higher. 

We can now test a posteriori the main assumption underlying our work, i.e. that 
early \VLII\ progenitors were populated 
with "pre-reionization" stellar systems. In the simulated high-resolution volume, there 
are 17,700 $M>10^6\, \msun$ subhalos at $z=11$, spread over a comoving volume of 
78 Mpc$^3$ containing a total mass of $M_{\rm DM}=3.5\times 10^{12}\,\msun$.      
Stars distributed according to a Salpeter IMF produce during their lifetime 
$f_\gamma\approx 4,000$ Lyman continuum 
photons per stellar proton, of which only a fraction $f_{\rm esc}$ will escape into the
intergalactic medium (IGM). The total stellar mass in these early-forming progenitors 
is $M_*=3.6\times 10^{10}\,\langle f_*\rangle \,\msun$. Hydrogen photoionization requires 
$(1+N_{\rm rec})$ photons above 13.6 eV
per atom, where $N_{\rm rec}$ is the number of radiative recombinations over 
a Hubble time. The total mass of intergalactic gas that can be kept ionized is 
then $M_{\rm ion}=f_\gamma M_* f_{\rm esc}/(1+N_{\rm rec})$. The condition
$M_{\rm ion}<(\Omega_b/\Omega_m)\,M_{\rm DM}$ then implies 
$\langle f_*\rangle <0.004\,[(1+N_{\rm rec})/f_{\rm esc}]$, where the factor in 
square brackets is of order 10 or so. The low star formation efficiencies derived in 
this work appear then to be consistent with the idea that the region around the Milky Way 
was reionized at $z\lta 11$ by external radiation, before stars in the Local Group  
formed in sufficient numbers (Weinmann \etal 2007). 

We conclude that star 
formation at first light must have occurred either with an IMF lacking stars 
below $0.9\,\msun$, or was intrinsically very inefficient in small dark matter halos. 
If the former, this may be an indication of an upward shift in the 
mass scale of the IMF at early cosmological times owing, e.g., to the hotter cosmic microwave background  
(Larson 2005). If the latter, our results may be 
viewed as another hint (see Strigari \etal 2008) of a minimum scale in galaxy 
formation, below which supernova feedback (Dekel \& Silk 1986) and/or H$_2$ 
photodissociation by a Lyman-Werner background (Haiman \etal 1997)   
sharply suppress star formation. 

\bigskip 
We thank J. Bullock and Z. Haiman for providing useful comments 
on an earlier draft. Support for this work was provided by NASA through
grants HST-AR-11268.01-A1 and NNX08AV68G (P.M.) and Hubble Fellowship
grant HST-HF-01194.01 (J.D.).  M.K.  gratefully acknowledges support
from the William L. Loughlin Fellowship at the Institute for Advanced
Study. 

{}

\end{document}